\documentclass[conference]{IEEEtran}

\usepackage{color}
\usepackage{cite}
\usepackage{url}
\usepackage{textcomp}
\usepackage{engord}
\usepackage{graphicx}
\usepackage{epstopdf}
\usepackage{algorithm}
\usepackage{algorithmicx}
\usepackage{algpseudocode}
\usepackage[cmex10]{amsmath}
\usepackage{amsfonts}
\usepackage{amssymb}
\usepackage{titlesec}
\usepackage{amsthm}
\usepackage{subfigure}
\usepackage{bm}
\usepackage{dsfont}
\usepackage{bbm}
\usepackage{booktabs}
\usepackage{multirow}
\theoremstyle{plain}

\usepackage[caption=false,font=normalsize,labelfont=sf,textfont=sf]{subfig}
\def\BibTeX{{\rm B\kern-.05em{\sc i\kern-.025em b}\kern-.08em
    T\kern-.1667em\lower.7ex\hbox{E}\kern-.125emX}}
    
\titleformat{\subsection}{\normalfont\itshape}{\thesubsection}{1em}{}
\titleformat{\subsubsection}{\normalfont\itshape}{\thesubsubsection}{1em}{}
\renewcommand{\thesubsection}{\Alph{subsection}.}
\renewcommand{\thesubsubsection}{\thesubsection\arabic{subsubsection}}

\begin{document}

\title{A Task Decomposition and Planning Framework for Efficient LLM Inference in AI-Enabled WiFi-Offload Networks}
\author{\IEEEauthorblockN{Mingqi Han\IEEEauthorrefmark{1}, Xinghua Sun\IEEEauthorrefmark{1} \\
\IEEEauthorblockA{\IEEEauthorrefmark{1}\small School of Electronics and Communication Engineering, Shenzhen Campus of Sun-Yat Sen University, Shenzhen, China\\
Email:hanmq@mail2.sysu.edu.cn, sunxinghua@mail.sysu.edu.cn}
\\
}
}
\maketitle
\IEEEpeerreviewmaketitle

\begin{abstract}
AI WiFi offload is emerging as a promising approach for providing large language model (LLM) services to resource-constrained wireless devices. However, unlike conventional edge computing, LLM inference over WiFi must jointly address heterogeneous model capabilities, wireless contention, uncertain task complexity, and semantic correlation among reasoning tasks. In this paper, we investigate LLM inference offloading in a multi-user multi-edge WiFi network, where each task can be executed locally, directly offloaded to a nearby edge access point (AP), or decomposed into multiple subtasks for collaborative execution across local and edge nodes. To this end, we propose a user-edge collaborative framework with an LLM-based planner that not only performs task decomposition but also infers subtask difficulty and expected output token length, enabling more accurate estimation of execution quality and latency on heterogeneous nodes. Based on these estimates, we further design a decomposition-aware scheduling strategy that jointly optimizes subtask assignment, execution, and aggregation under communication, queuing, and computation constraints. Simulation results show that the proposed framework achieves a better latency-accuracy tradeoff than local-only and nearest-edge baselines, reducing the average latency by $20\%$ and improving the overall reward by $80\%$. Moreover, the distilled lightweight planner approaches the performance of the large teacher model while remaining more suitable for practical edge deployment.
\end{abstract}


\section{Introduction}
Recent advances in large language models (LLMs) have significantly expanded wireless intelligent services, including multi-modal perception, interactive reasoning, and agentic task execution. Meanwhile, home, enterprise, and industrial wireless networks are increasingly populated by client devices such as cameras, robots, wearables, and lightweight terminals that require AI functionality but often lack sufficient local computing resources due to constraints on chip capability, power budget, cost, and thermal design. As a result, moving AI inference closer to users through wireless-connected edge platforms is becoming a practical and increasingly important design direction \cite{10685369}.

\par This trend is also reflected in ongoing standardization discussions on AI offloading over IEEE 802.11 networks. Recent IEEE 802.11 discussions on AI offloading highlight an emerging Compute-as-a-Service \cite{devegt2026aioffload}, in which a WiFi access point (AP) or other WiFi-enabled edge device advertises AI capabilities, accepts inference requests from constrained clients, and returns inference results. The envisioned service workflow includes capability discovery, session setup, model selection or loading, inference input/output exchange, and session tear-down, and is motivated by practical requirements such as low latency, privacy preservation, reliability, and interoperability across devices and vendors.

\par From the perspective of AI service provisioning, AI WiFi offload is fundamentally different from conventional edge computing. First, the workload is no longer a traditional computation task with explicit resource demands, but rather an LLM inference task whose complexity is hard to characterize in advance. Since LLM tasks span heterogeneous forms such as question answering, reasoning, multi-modal understanding, and agentic interaction, it is generally difficult to accurately quantify the task difficulty, output size, and expected inference quality before execution. Second, the execution cost of LLM inference cannot be captured by a single CPU-cycle or computation-capacity abstraction widely used in conventional edge computing \cite{9789202, 11017483, 11125971}. Instead, it jointly depends on multiple model- and hardware-related factors, including GPU computing capability, memory bandwidth, model size, and model type, all of which directly affect the latency and response quality of offloading decisions. Third, traditional edge computing mainly focuses on binary offloading \cite{9789202, 11017483} and partial offloading \cite{11125971}. Binary offloading typically treats a task as an indivisible unit and therefore cannot exploit the reasoning of many LLM inference tasks, e.g., chain-of-thought and graph-of-thought \cite{huang2024understanding}. On the other hand, partial offloading usually assumes continuously divisible tasks and thus fails to capture the semantic dependency and logical correlation among language subtasks. Moreover, neither scheme explicitly considers that a suitable decomposition strategy may itself improve inference quality by matching subtasks to heterogeneous models with different strengths. Therefore, AI WiFi offload requires a new formulation that jointly considers wireless contention, heterogeneous model characteristics, decomposition overhead, semantic dependency among subtasks, and the impact of collaborative execution on final answer quality.

\par Recent studies have also begun to explore collaborative LLM inference across local devices and edge nodes. \cite{dong2025hybridflow} proposes an edge-cloud collaborative inference framework that constructs a dependency-aware graph for each query and dynamically routes unlocked subtasks between edge and cloud, thereby improving both the inference cost and response accuracy. Meanwhile, \cite{lv2025collaboration} proposes a local-cloud collaboration paradigm in which a cloud-hosted large-scale LLM and an on-device lightweight model cooperate during inference. Specifically, the cloud-hosted LLM generates high-quality candidate results while the model deployed on local device refines them using the local information. These studies provide useful evidence that collaborative inference and task-level coordination can improve efficiency and service quality beyond pure local or pure remote execution. However, they still differ substantially from the AI WiFi-offload scenario considered in this paper. Existing works typically focus on device--cloud or edge--cloud cooperation for a single service pipeline, without explicitly modeling the multi-user multi-edge setting where multiple UEs compete for wireless access and heterogeneous edge nodes simultaneously provide inference services. In addition, they generally abstract communication as a high-level delay or budget term, rather than incorporating practical WiFi protocol behaviors such as contention, handshaking, and inter-node coordination overhead. More importantly, these studies do not investigate collaborative inference under realistic WiFi offload mechanisms, nor do they fully address how decomposition, semantic dependency among subtasks, and heterogeneous model capabilities jointly affect both latency and final inference quality in distributed local-edge execution.

\par Motivated by this, we optimize the LLM inference offloading in a multi-user multi-edge WiFi network, where multiple user equipments (UEs) compete for wireless resources while heterogeneous edge APs provide different LLM services. In the considered system, each UE is equipped with a lightweight local model and can either execute a task locally, offload the entire task to an edge AP, or decompose the task into several subtasks for collaborative execution across the local device and multiple edge nodes. Through collaboration among small-scale LLM deployed on APs and UEs, such a design can alleviate local computational limitations and improve response quality.

\par However, enabling efficient AI WiFi-offload remains challenging. Offloading to edge APs can improve answer quality and reduce the computation burden on resource-constrained UEs by leveraging stronger models, but it also introduces additional communication and coordination overhead. In particular, multi-node collaborative execution incurs uplink transmission delay, WiFi contention, subtask dispatch among edge nodes, result collection, and final answer aggregation. Moreover, although larger edge models generally provide stronger reasoning capability, they may also suffer from heavier queuing delay under multi-user traffic. As a result, a practical AI WiFi-offload design must jointly balance communication delay, computation delay, queuing delay, and inference quality. This tradeoff becomes even more subtle when task decomposition is optional: while suitable decomposition can better exploit heterogeneous node capabilities and improve execution flexibility, unnecessary splitting may offset the gain because of extra coordination overhead.

\par To address these challenges, this paper develops an LLM-based task decomposition and planning framework for AI WiFi-offload networks. Specifically, a planner LLM is employed not only to determine whether an incoming task should be decomposed, but also to infer key subtask attributes during the decomposition process, including subtask difficulty and the expected output token length. Based on these inferred attributes, the system can further estimate the execution accuracy and latency of each subtask on different candidate edge APs and the local device by considering heterogeneous model capability, computing resources, and communication overhead. With such task-aware and node-aware information, the planner is able to perform more informed subtask assignment, scheduling, and aggregation under the network objective function, thereby jointly improving inference quality and end-to-end latency for LLM services in WiFi offload networks. The main contributions of this paper are summarized as follows:
\begin{itemize}
    \item We propose a UE-AP collaborative framework for LLM inference in WiFi networks, where resource-constrained UEs and heterogeneous edge APs jointly execute reasoning tasks through decomposition and planning. Through this framework, both the execution latency and response performance can be enhanced.
    \item We develop an LLM-based task decomposition and planning method that not only splits the original task into executable subtasks but also infers subtask difficulty and output length, which are further used to estimate the execution accuracy and latency on different nodes for subtask scheduling and aggregation. To enable practical deployment at the edge, we further distill planning capability of a teacher model into a lightweight planner.
    \item Simulation results show that the proposed task decomposition and planning framework outperforms other baselines, reducing the average latency by $20\%$ and improving the overall reward by $80\%$ compared with offloading to the nearest edge AP. Moreover, the distilled lightweight planner approaches the performance of the large-scale teacher LLM, achieving comparable reward with even lower latency, which demonstrates its effectiveness for practical edge deployment.
\end{itemize}

\section{System Model}
We consider a WiFi edge network with a set of UEs $\mathcal{U}=\{1,\dots,U\}$ and a set of edge APs $\mathcal{E}=\{1,\dots,E\}$.  The position of UE $u\in\mathcal{U}$ and edge AP $e\in\mathcal{E}$ is denoted as ${\bf p}_u$ and ${\bf p}_e$, respectively. Each edge AP is independently equipped with a GPU and LLM model, in which the computing capability is denoted as $F_e$ (in TFLOPS), the memory bandwidth is denoted as $A_{bm, e}$ and the model size is denoted as $P_e$. Similarly, UE $u$ has local capability $F_u$, local GPU memory bandwidth $A_{bm, u}$ and local model size $P_u$.

\par In this scenario, we consider that only UEs have LLM task requirements, in which task generation follows a Poisson arrival and the arrival rate is $\lambda_u$. In particular, new task generates only when all previous tasks complete. For a task generated by UE $u$, the system supports three execution modes:
\begin{enumerate}
    \item \textbf{Local-only:} the task is solved entirely on the UE.
    \item \textbf{Offload-edge:} the task is offloaded to the nearest AP and solved without decomposition.
    \item \textbf{Decomposition and Plan:} the task is offloaded to the planner and optionally decomposed into executable subtasks. Then the subtasks are assigned to local UE or edge APs for collaborative inference and final aggregation. It is assumed that the planner is executed at the nearest edge AP of the requesting UE.
\end{enumerate}

\noindent In particular, each UE can only start a new task after its current task finishes.

\subsection{Task and Decomposition Model}
The LLM inference task can be characterized as
\begin{equation}
    \tau = (\text{id},u,s,c,p,r),
\end{equation}

\noindent in which $u$ is the source UE, $s$ is the task source, $c$ is the task category, $p$ is the prompt, and $r$ is the reference answer.

\par In the decomposition mode, the task is decomposed into multiple subtasks through the planner
\begin{equation}
    \mathcal{S}(\tau)=\{s_1,s_2,\dots,s_K\},
\end{equation}
where each subtask contains a role, task type, prompt, offload edge AP, aggregator edge AP, and a dependency set. In particular, the planner is instructed to return a single subtask when splitting is unnecessary, i.e., $K=1$. Here, the dependency set only represents the relationship between subtasks instead of the sequential execution. After decomposition, all subtasks are transmitted to the corresponding edge APs selected by the planner for parallel execution. After execution, each subtask is transmitted to the selected aggregator edge AP to obtain the final answer, which is then transmitted to the source UE. In the following, we first illustrate the communication model for transmission, and then the LLM inference model.

\subsection{WiFi Communication Model}
\begin{figure}[t]
\centering
\includegraphics[width=82mm,height=32mm]{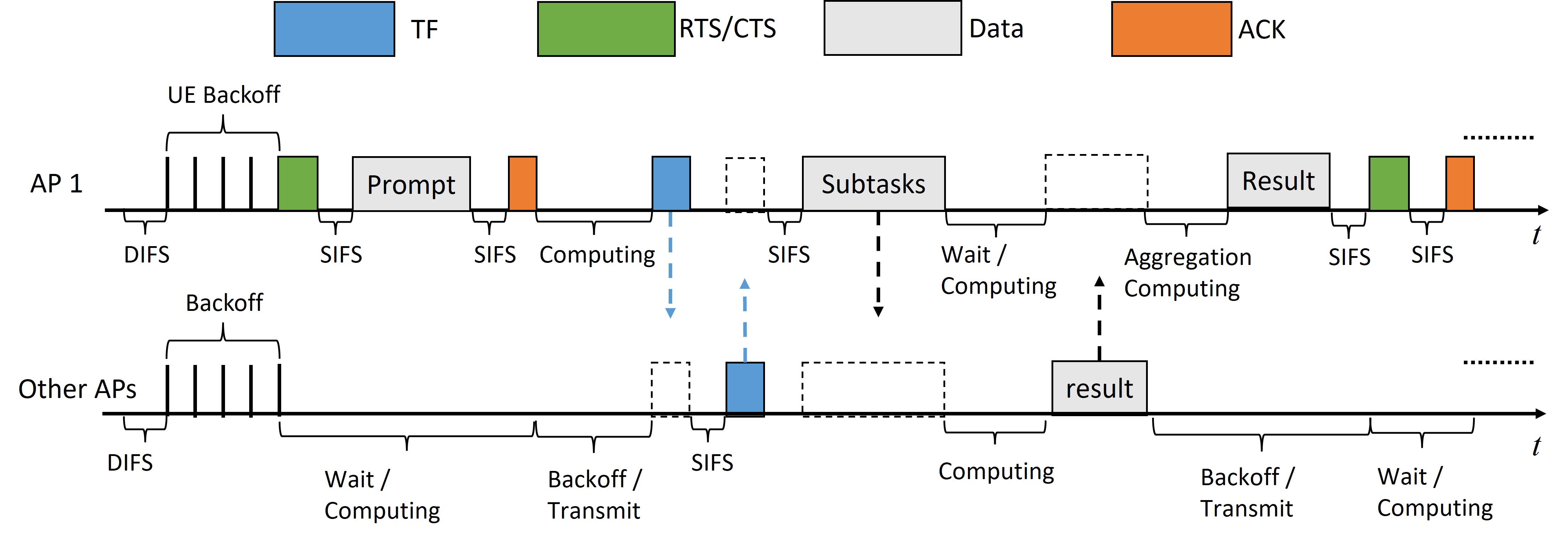}
\caption{Transmission and computing procedure for edge APs in this scenario.}
\label{fig:access-procedure}
\end{figure}

\par In this paper, we adopt a WiFi 7 inspired communication model including binary exponential backoff, RTS/CTS handshaking, acknowledgment (ACK) overhead, and trigger-frame (TF) based inter-edge exchange. The channel bandwidth is $B$, while UE and edge transmit powers are denoted by $P_{tx, u}$ and $P_{tx, e}$, respectively.

\par The overall access procedure is illustrated in Fig. \ref{fig:access-procedure}. At the beginning, each UE starts backoff when sensing the channel for a period of DIFS. The UE that completes backoff first uploads the task prompt to the nearest edge AP through a RTS/CTS mechanism. In particular, when multiple UEs complete backoff simultaneously, both transmissions collide, increase the contention window and restart backoff. Then, the edge AP decomposes the original task into subtasks through the planner model. Then, it broadcasts the TF to other APs and the source UE to collect information for planning, including queuing, LLM model status, computing capability and memory bandwidth. Accordingly, it can plan the offload edge AP for each subtask in a greedy manner, which will be detailed in the next section. Subsequently, these APs calculate the received subtasks, and transmit the sub-results to the aggregator. Finally, the aggregator edge AP generates the final result and transmits it to the source UE. In the following, we analyze the overall communication latency.

\par For a packet of size $L$ bits transmitted over a distance $d$, the data transmission delay is
\begin{equation}
    T_{\mathrm{tx}}(L,d)=\frac{L}{R(d)},
\end{equation}
in which the transmission rate $R(d)$ is
\begin{equation}
R(d)=B\log_2(1+\frac{\xi P_{tx}G(d)}{N_0B}) .
\end{equation}
\noindent in which the path gain is given by
\begin{equation}
    G(d)=10^{-PL(d)/10} ,
\end{equation}
Here, $\xi$ is the small-scale Rayleigh fading, $\mathrm{PL}(d)$ is the path loss in dB, and $N_0$ is the noise power spectral density. In particular, we introduce the classical TGax model for indoor WiFi scenarios \cite{TGax}
\begin{equation}\label{equ:path-loss}
\mathrm{PL}(d)=(40.05+20 \log _{10}\left(\frac{\min \left(d, B_p\right) f_c}{2.4}\right)+P^{\prime}+7 W_n) ,
\end{equation}

\noindent in which $d$ is the distance between the transmitter and the receiver in meters. $f_c$ denotes the central frequency in GHz, $W_n$ is the number of walls, and $B_p$ denotes the breaking point of the path loss. $P^{\prime}$ is given by
\begin{equation}
P^{\prime} =  \begin{cases}
0, & \text { if } d \leq B_p \\
35 \log _{10}\left(d / B_p\right), & \text { if } d \textgreater B_p
\end{cases} .
\end{equation}

\par In this paper, since we focus on the LLM inference task, we introduce the expected holding time and steady probability for request $\tau_{R_i}$, $\pi_{R_i}$ to capture WiFi contention for other conventional messages \cite{6212505}, which increases with respect to the number of APs and UEs. The expected CSMA/CA contention delay is defined as
\begin{equation}
    T_{\mathrm{cont}}=T_{\mathrm{DIFS}}+\sum\limits_{i=1}^C \pi_{R_i} \cdot \tau_{R_i},
\end{equation}
where $\mathrm{CW}_{\min}$ is the initial contention window, $C$ is the cutoff phase and $R_i$ represents the requesting phase for backoff stage $i$. The uplink transmission delay from UE to AP with prompt length $L$ and distance $d$ is then modeled as
\begin{equation}
\begin{aligned}
    T_{u,e}^{\mathrm{UL}}(L,d)=&\;T_{\mathrm{cont}}+T_{\mathrm{RTS}}+T_{\mathrm{SIFS}}+T_{\mathrm{CTS}} \\
    &+T_{\mathrm{SIFS}}+T_{\mathrm{tx}}(L,d)+T_{\mathrm{SIFS}}+T_{\mathrm{ACK}}.
\end{aligned}
\end{equation}

\noindent Similarly, other communication parts including inter-edge data transmission and result return, are modeled via TF notification plus data exchange. The simulator separately accumulates: 1) UE-to-orchestrator upload $T_{u,e}^{\mathrm{UL}}(L,d)$, 2) planner-to-edge subtasks transmission $T_{e1,e2}^{\mathrm{P2E}}(L,d)$, 3) edge-to-aggregator result transmission $T_{e1,e2}^{\mathrm{E2A}}(L,d)$, and 4) final downlink from aggregator to source UE $T_{e,u}^{\mathrm{DL}}(L,d)$. Accordingly, the overall communication latency is given by 
\begin{equation}
\begin{aligned}
    T_\tau^\text{comm}=&T_{u,e}^{\mathrm{UL}}(L,d) + \max\limits_{e2} T_{e1,e2}^{\mathrm{P2E}}(L,d)  \\
    & +  \max\limits_{e1}T_{e1,e2}^{\mathrm{E2A}}(L,d) + T_{e,u}^{\mathrm{DL}}(L,d) .
\end{aligned}   
\end{equation} 

\noindent in which $\max\limits_{e2} T_{e1,e2}^{\mathrm{P2E}}(L,d)$ and $\max\limits_{e1}T_{e1,e2}^{\mathrm{E2A}}(L,d)$ takes the maximum transmission time for parallel subtask transmission, and other terms are executed in a sequential manner.

\subsection{Inference Latency Model}
LLM inference consists of a prefill phase and a decode phase: the prefill phase processes all prompt tokens in parallel and is mainly compute-bound, while the decode phase generates tokens one by one and is mainly limited by GPU memory bandwidth \cite{yuan2024llm}. Since the forward computation per token is approximately proportional to $2P$, the prefill latency can be approximated as
\begin{equation}
T_{\mathrm{prefill}} \approx \frac{2 b s P}{F},
\label{eq:prefill_latency}
\end{equation}

\noindent in which $P$ denotes the size of model parameters, $s$ is the input prompt length, $F$ is the GPU computing capability and $b$ is the batch size. During decoding, the dominant overhead comes from repeatedly loading model weights and KV-cache data from memory. Thus, the per-token decode latency can be approximated as
\begin{equation}
T_{\mathrm{decode}} \approx \frac{2 P}{A_{bm}},
\label{eq:decode_latency}
\end{equation}

\noindent in which $A_{bm}$ denotes the GPU memory bandwidth. Accordingly, the total inference latency for generating $o$ tokens is
\begin{equation}
T_\tau^{\mathrm{cmp}} \approx T_{\mathrm{prefill}} + o\, T_{\mathrm{decode}}.
\label{eq:total_latency}
\end{equation}

\noindent Similarly, the inference latency for aggregating $K$ subtasks is 
\begin{equation}
T_\tau^{\mathrm{agg}} \approx \sum\limits_{k=1}^K T_{\mathrm{prefill}}^k + o\, T_{\mathrm{decode}}.
\end{equation}
\noindent in which $T_{\mathrm{prefill}}^k$ is the prefill time for $k$-th subtask and $o$ is the output length of the aggregator.

\subsection{Objective Model}
For non-decomposed execution, the total latency is
\begin{equation}
    T_{\tau}^{\mathrm{single}} = T_{\tau}^{\mathrm{cmp}} + T_{\tau}^{\mathrm{comm}} + T_{\tau}^{\mathrm{que}}.
\end{equation}

For decomposed collaborative execution, the task latency is given by
\begin{equation}
\small
    T_{\tau}^{\mathrm{prop}} = T_{\tau}^{\mathrm{cmp}, p} + \max\limits_{k} (T_{\tau}^{\mathrm{cmp}, k} + T_{\tau}^{\mathrm{que}, k}) + T_{\tau}^{\mathrm{agg}} + T_{\tau}^{\mathrm{comm}},
\end{equation}
where $T_{\tau}^{\mathrm{cmp}, p}$ is the computing latency for planning, $\max\limits_{k} (T_{\tau}^{\mathrm{cmp}, k} + T_{\tau}^{\mathrm{que}, k})$ is the waiting time to complete all $K$ subtasks, $T_{\tau}^{\mathrm{agg}}$ is the aggregation latency, and $T_{\tau}^{\mathrm{comm}}$ includes all WiFi exchanges among UE, orchestrator AP, worker nodes, and aggregator. 

\par Finally, considering both the task latency and the response accuracy, the objective is presented as follows:
\begin{equation}\label{equ:optimization-problem}
	 \max \frac{1}{|\mathcal T_{\mathrm{done}}|}\sum_{\tau\in\mathcal T_{\mathrm{done}}}
\left[\omega_a\mathbbm{1}(\hat r_\tau=r_\tau)-\omega_dT_\tau\right]
\end{equation}

\noindent in which $\mathcal T_{\mathrm{done}}$ is the set of all completed tasks. $\omega_d$ and $\omega_a$ are weights for average delay and response accuracy, respectively. In particular, $r_\tau$ is only utilized to evaluate the correctness of generated responses $\hat r_\tau$, which cannot be obtained during scheduling.

\section{Plan Algorithm}
In this section, we illustrate the proposed task decomposition and planning algorithm. In the following, we first illustrate the planning strategy, the planner training, and then the algorithm overview.

\subsection{Plan Strategy}
To jointly account for latency and answer quality, we employ a planner to estimate subtask characteristics prior to execution. Specifically, for each subtask $s_k$, the planner provides: 1) an estimate of the expected output token length, and 2) an estimate of the achievable answer correctness when executed on different nodes.

\textbf{Computation latency estimation:} The planner first predicts the expected number of generated tokens $\hat{n}_g$ for each subtask based on its prompt and task type. Given the prompt length $n_p$ and the estimated output length $\hat{n}_g$, the computation latency is evaluated using (\ref{eq:total_latency}), i.e.,
\begin{equation}
    T_{k,i}^{\mathrm{cmp}} = T^{\mathrm{cmp}}(n_p,\hat{n}_g,M_i,F_i, A_{bm}),
\end{equation}
which depends on the model size $M_i$ and the computing capability $F_i$ of node $i$.

\textbf{Correctness estimation:} The planner also estimates the expected correctness $\hat{a}_{k,i}$ for executing subtask $s_k$ on node $i$ based on the task content, difficulty, and model capability. This estimate implicitly captures differences across model scales and task types.

Based on these estimates, the overall scheduling score is defined as
\begin{equation}\label{equ:score}
    J_{k,i}= \omega_a \hat{a}_{k,i} - \omega_d T_{k,i}^{\mathrm{tot}},
\end{equation}
in which the estimated latency $T_{k,i}^{\mathrm{tot}}$ is given by
\begin{equation}
    T_{k,i}^{\mathrm{tot}} = T_{k,i}^{\mathrm{tx}} + T_i^{\mathrm{que}} + T_{k,i}^{\mathrm{cmp}}.
\end{equation}

\par According to (\ref{equ:score}), the node with the highest score is selected as the execution destination for the subtask $k$.

\subsection{Planner Training}
Due to limited computational capacity at edge APs, directly deploying large-scale LLMs for planning is impractical. This leads to two issues: 1) insufficient task decomposition capability for small-scale LLMs, and 2) inaccurate estimation of subtask attributes due to mismatch between planning and pre-training data. To address this, we fine-tune a lightweight planner using supervision from a stronger teacher LLM. Specifically, the DeepSeek-v3.2 (DS-v3.2) is used to generate high-quality planning outputs, while the Qwen2.5-7B-Instruct model serves as the student model deployed at the edge.

\par Given input task $\mathbf{x}$ and teacher output $\mathbf{y}^\ast$, the student model is trained using supervised fine-tuning with the following loss:
\begin{equation}
    \mathcal{L}_{\mathrm{SFT}}(\theta)
    =
    - \mathbb{E}_{(\mathbf{x},\mathbf{y}^\ast)}
    \left[
        \sum_{t=1}^{T}
        \log P_{\theta}(y_t^\ast \mid \mathbf{x}, y_{<t}^\ast)
    \right].
\end{equation}

\noindent This enables the edge-deployed planner to learn effective task decomposition and estimation capabilities under resource constraints.

\begin{figure*}[t]
	\centering
	\subfigure[Task Accuracy]{
	\includegraphics[width=56mm,height=40mm]{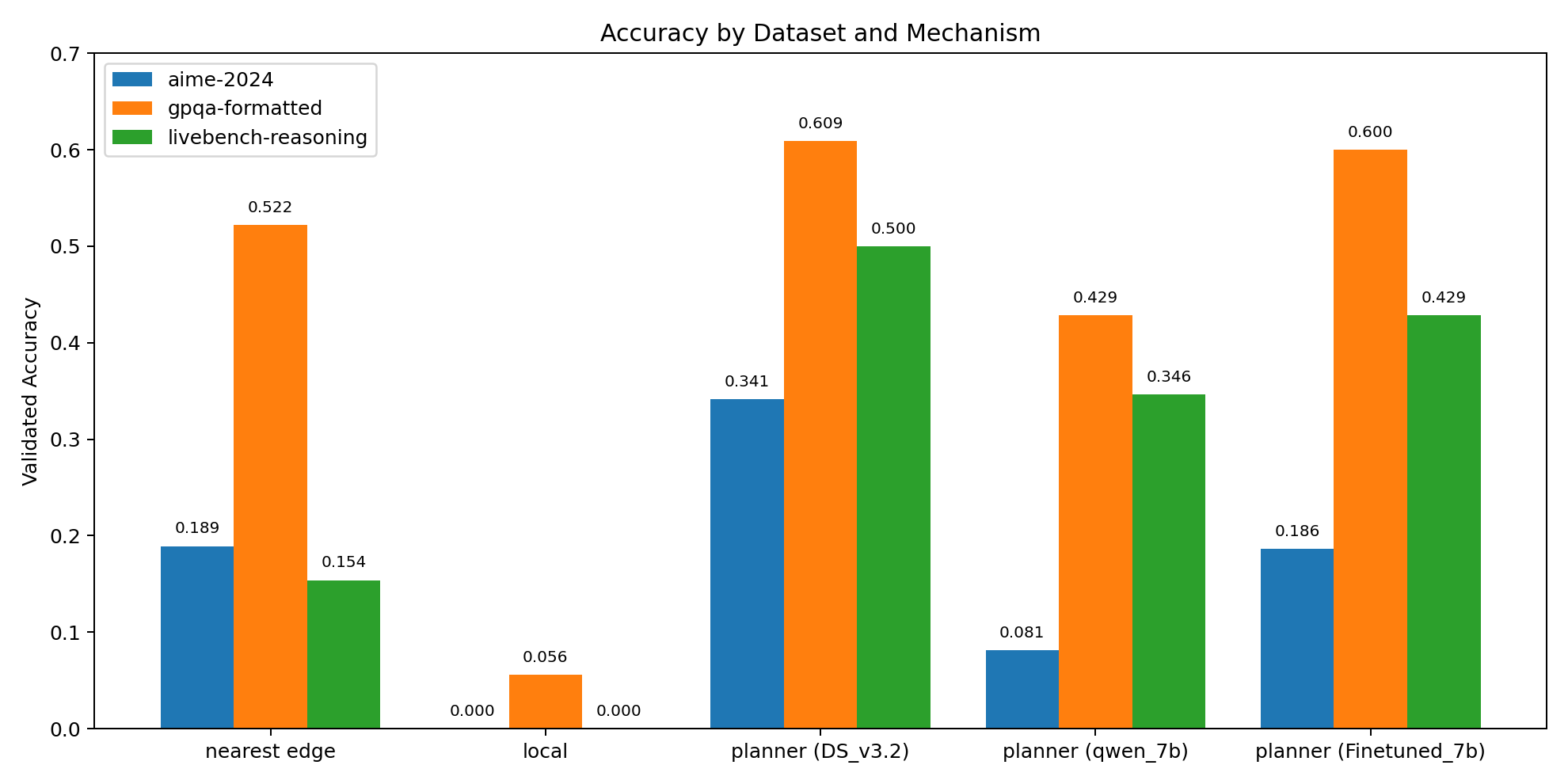}}
	\subfigure[Average latency]{
	\includegraphics[width=56mm,height=40mm]{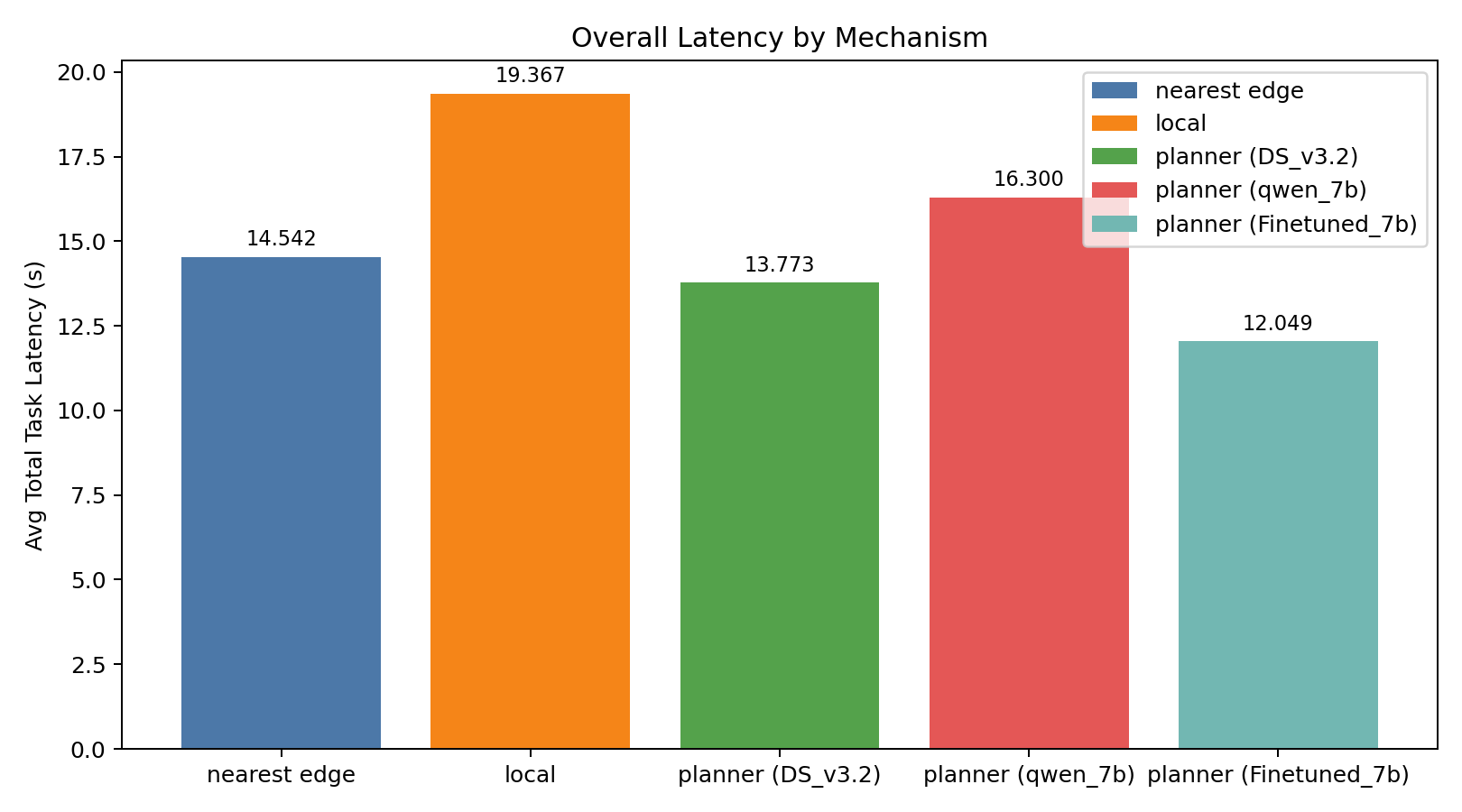}}
	\subfigure[Overall Reward]{
	\includegraphics[width=56mm,height=40mm]{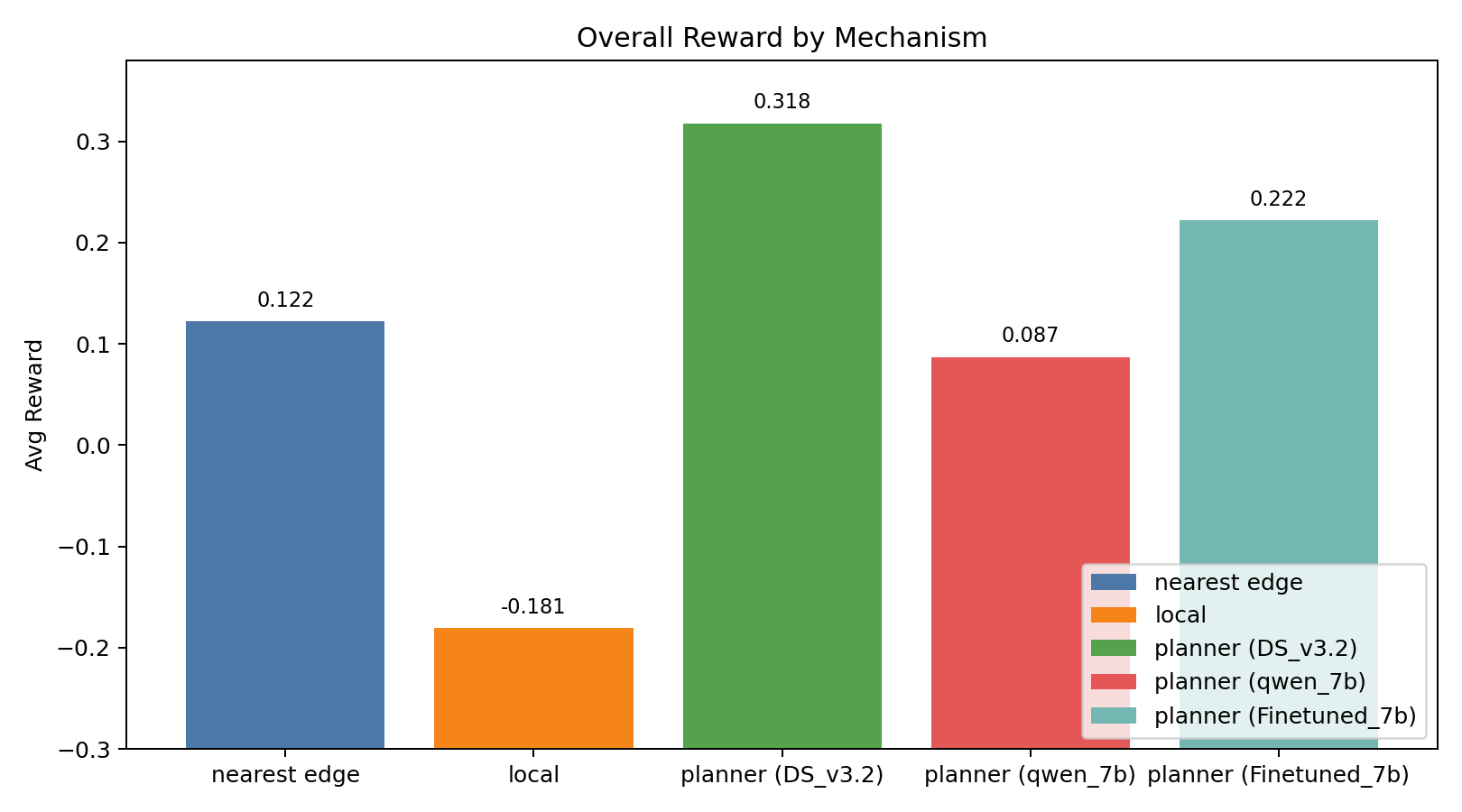}}
	\caption{Performance comparison between approaches and task types in AI-enabled WiFi-Offload Networks.}
	\label{fig:all_simulations}
\end{figure*}

\subsection{Algorithm Overview}
The proposed framework integrates planner-guided task decomposition with decomposition-aware scheduling. Given an incoming task, the nearest edge AP first invokes the planner to determine whether decomposition is beneficial and to generate a structured subtask set. Then, the planner assigns each subtask to an appropriate execution node and decide the aggregator.

\par For each subtask, the planner evaluates all candidate nodes, including local UE and edge APs, using a unified cost function (\ref{equ:score}) that combines communication delay, queuing delay, computation delay, and planner-estimated correctness. Then, the node with the highest weighted score is selected to execute this subtask. After all subtasks are executed, they are gathered in an aggregation node to produce the final answer. Algorithm~\ref{alg:proposed} summarizes the overall procedure.

\begin{algorithm}[t]
\small
\caption{Task Decomposition and Planning}
\label{alg:proposed}
\begin{algorithmic}[1]
\Require UE $u$, task $\tau$, edge AP set $\mathcal{E}$, weights $\omega_d,\omega_a$
\Ensure Final answer for task $\tau$
\State UE $u$ uploads task $\tau$ to the nearest edge AP $e$
\State Planner at $e$ generates subtask set $\mathcal{S}(\tau)$, estimates subtask attributes, and determines the aggregator node
\If{$|\mathcal{S}(\tau)|=1$}
    \State Execute the task locally or at the selected edge AP
    \State \Return final answer
\EndIf
\State Initialize subtask assignment
\For{each subtask $s_k\in \mathcal{S}(\tau)$}
    \For{each candidate node $i\in \{u\}\cup\mathcal{E}$}
        \State Estimate $T_{k,i}^{\mathrm{tx}},\; T_i^{\mathrm{que}},\; T_{k,i}^{\mathrm{cmp}},\; \hat{a}_{k,i}$
        \State Compute score $J_{k,i}$ according to (\ref{equ:score})
    \EndFor
    \State Assign subtask $s_k$ to node $i$ with the highest $J_{k,i}$ score for execution
    \State Update node queues and subtask states
\EndFor
\State Assigned nodes transmit subtask results to the aggregator
\State Aggregate subtask outputs and return the final answer
\end{algorithmic}
\end{algorithm}

\section{Simulation Results}
\subsection{Simulation Setup}
We evaluate the proposed framework using the following simulation environment. We consider a typical $50 m\times 50m$ indoor environment with multiple edge APs and UEs. The number of edge APs varies, with each AP serving a different number of UEs within its coverage area. In each episode, both the number and position of APs and UEs are randomly generated. All simulation results are averaged by $10$ independent episodes, with each episode simulating $600$ seconds. To ensure a wide range of LLM inference, tasks are sampled from three sources: AIME-2024 (Math), LiveBench-Reasoning (Daily), and GPQA (Science). The detailed network parameter is summarized in Table \uppercase\expandafter{\romannumeral1}.

For comprehensive evaluation, the following methods are introduced for comparison.
\begin{itemize}
    \item \textbf{Proposed Planner}: decomposes the original task into subtasks and plans the offload nodes and aggregator.
    \item \textbf{Nearest-Edge}: the whole task is offloaded to the nearest AP without decomposition.
    \item \textbf{Local-Only}: the whole task is solved on the UE without decomposition.
\end{itemize}

\begin{table}[t]
\centering
\caption{Main Simulation Parameters}
\label{tab:params}
\begin{tabular}{lc}
\toprule
Parameter & Value \\
\midrule
WiFi bandwidth & 40 MHz \\
Transmission power $P_{tx, u}$, $P_{tx, e}$ & 17, 25 dBm \\
Noise density $N_0$ & $-169$ dBm/Hz \\
Central frequency $f_c$ & $5$ GHz \\
Number of edge APs $E$ & $\left[2, 4\right]$ \\
Edge model sizes & 7B / 14B / 32B \\
Edge computing capability $F_e$ & $\left[120, 312 \right]$ TFLOPS (fp16) \\
Edge memory bandwidth $A_{bm,e}$ & $\left[0.6, 2 \right] \times 10^{12}$ Bytes/s \\
Number of UEs $U$ & $\left[5, 7\right]$ \\
UE local model size & 1.5B \\
UE computing capability $F_u$ & $\left[20, 48 \right]$ TFLOPS (fp16) \\
UE memory bandwidth $A_{bm,u}$ & $\left[0.2, 0.4 \right] \times 10^{12}$ Bytes/s \\
Task arrival rate $\lambda_u$ & $0.1\,\mathrm{s}^{-1}$ \\
Delay weight $\omega_d$ & 0.01 \\
Correctness weight $\omega_a$ & 1.0 \\
Indoor walls $W_n$ & 2 \\
TGax breaking point $B_p$ & 10 m \\
Maximum number of subtasks $K$ & 3 \\
\bottomrule
\end{tabular}
\end{table}

\subsection{Performance}
The performance comparison is summarized in Fig.~\ref{fig:all_simulations}. In terms of overall reward, the DS-v3.2 planner achieves the best result ($0.318$), followed by the proposed fine-tuned planner ($0.222$), and both clearly outperform Nearest-Edge ($0.122$) and Local-Only ($-0.181$). In terms of latency, the proposed fine-tuned planner attains the lowest average delay ($12.049$s), which is lower than that of DS-v3.2 ($13.773$s) and Nearest-Edge ($14.542$s), whereas Local-Only incurs a higher delay of $19.367$s. By contrast, the original Qwen-7B planner yields both higher latency ($16.300$s) and lower reward ($0.087$), revealing its limited planning effectiveness. The dataset-wise accuracy results further confirm that DS-v3.2 consistently provides the strongest performance, while the proposed fine-tuned planner remains very close to it, e.g., $0.600$ versus $0.609$ on GPQA and $0.429$ versus $0.500$ on LiveBench-Reasoning. In contrast, both Nearest-Edge and Local-Only perform much worse, especially on more challenging tasks such as AIME, where their accuracies are only $0.189$ and $0$, respectively.

\par These results demonstrate that effective task decomposition and accurate planning are critical to improving the latency--accuracy tradeoff in AI WiFi-offload systems. The Local-Only scheme is fundamentally constrained by the limited computing capability of UEs, leading to both excessive latency and poor overall reward. Although Nearest-Edge reduces latency by offloading the entire task to the closest AP, it cannot fully exploit heterogeneous edge resources or adapt execution to different task characteristics. DS-v3.2 provides a strong teacher baseline by generating high-quality decompositions and more reliable execution estimates, but its model scale makes direct deployment at the edge impractical. In contrast, the original Qwen-7B planner lacks sufficient decomposition and estimation capability, which leads to inefficient collaboration and unnecessary communication and queuing overhead. Benefiting from teacher-guided fine-tuning, the proposed planner effectively inherits stronger decomposition and estimation ability while preserving the deployment advantage of a lightweight model, thereby achieving a more desirable balance between accuracy and latency for practical edge implementation.

\section{Conclusion}
This paper investigated LLM inference offloading in a multi-UE multi-edge WiFi network and developed a unified framework that jointly captures optional task decomposition, WiFi communication overhead, heterogeneous inference latency, and distributed result aggregation. Based on this framework, we proposed a decomposition-aware weighted scheduling algorithm that incorporates communication delay, queuing delay, computation delay, and estimated correctness into the offloading decision. Simulation results showed that simply executing tasks locally or directly offloading them to the nearest edge node cannot effectively balance latency and answer quality, whereas decomposition-aware collaborative execution can better exploit heterogeneous edge resources and achieve a more favorable performance tradeoff. The results also verified that a lightweight planner distilled from a stronger teacher model can approach the performance of a large planner while remaining more suitable for practical edge deployment. Overall, this work demonstrates the potential of planner-guided collaborative inference for future AI-enabled WiFi-offload systems and provides a useful basis for further studies on adaptive decomposition, more realistic MAC-layer modeling, and joint communication-computation orchestration.

\bibliographystyle{IEEEtran}
\bibliography{ref}

\end{document}